\documentclass[10pt,a4paper,twocolumn,english,pra,aps,showpacs,floatfix,groupedaddress,superscriptaddress]{revtex4}

\usepackage{graphicx}
\usepackage{epsfig}
\usepackage[english]{babel}
\usepackage{amsmath}
\usepackage{amssymb}
\usepackage{amsfonts}
\usepackage{longtable}
\usepackage{dcolumn}
\usepackage{bm}
\usepackage{bbm}

\begin{document}
\title{Generalization of short coherent control pulses: 
extension to arbitrary rotations}

\author{S. Pasini}
\email{pasini@fkt.physik.tu-dortmund.de}
\affiliation{Lehrstuhl f\"{u}r Theoretische Physik I, Dortmund University of 
Technology, Otto-Hahn Stra\ss{}e 4, 44221 Dortmund, Germany}
\author{G.S. Uhrig}
\email{goetz.uhrig@tu-dortmund.de}
\affiliation{Lehrstuhl f\"{u}r Theoretische Physik I, Dortmund University of 
Technology, Otto-Hahn Stra\ss{}e 4, 44221 Dortmund, Germany}

\date{\rm\today}

\begin{abstract}
We generalize the problem of the coherent control of small quantum systems to 
the case where the quantum bit (qubit) is subject to a fully general rotation. 
Following the ideas developed in Pasini \textit{et al} (2008 \textit{Phys.~Rev.}~A \textbf{77}, 032315), 
the systematic expansion in the shortness of the pulse
is extended to the case where the pulse acts on the qubit as a 
general rotation around an axis of rotation varying in time.  The 
leading and the next-leading corrections are computed. For certain pulses we
prove that the general rotation does not improve on the simpler rotation
with fixed axis.
\end{abstract}


\maketitle

\section{Introduction} 
The ability to maintain a two-level system, for instance a $S=1/2$ spin or
a quantum bit (qubit) in the language of quantum information, in a coherent 
state as long as possible has always been 
of vital importance in nuclear magnetic resonance (NMR). Nowadays, 
decoherence is considered one of the basic difficulties to overcome 
\cite{zolle05} for the realization of a quantum information devices. The 
decoherence of a qubit, i.e., its decay  from an initial state to
a mixture, is attributed to the coupling of the qubit to the 
macroscopic environment.

Among the numerous techniques developed to cope with decoherence 
we consider the dynamical decoupling (DD) \cite{viola98,ban98,vande04}. 
The central idea of DD is to disentangle the qubit from 
the bath by means of repetitive, instantaneous rotations in spin space
that prevent the qubit from precessing. Put sloppily, the coupling to
the environment is averaged to zero.
It has been shown that various optimizations of such sequences are possible. 
We cite here the Carr-Purcell and 
Meiboom-Gill sequence \cite{carr54,meibo58,haebe76}, the concatenated 
sequence proposed by Khodjasteh and Lidar \cite{khodj05,khodj07}, and the fully
optimized non-equidistant (UDD) sequence \cite{uhrig07,uhrig08}. 

The necessary rotations for DD are achieved by means of ideal $\pi$ pulses
which are instantaneous and infinitely strong, i.e., $\delta$ peaks. 
During  their application the qubit has no time to experience the effect 
of the bath. Thus the rotation of the qubit can be treated separated from 
the evolution due to the qubit-plus-bath Hamiltonian.
Note that ideal pulses on single qubits correspond to single qubit gates
in the framework of quantum information processing.

Given the significant interest in coherent control by short pulses
there have been already a number of previous investigations
of the effects of the realistically finite pulse lengths $\tau_p >0$.
For instance, their cumulative effect in pulse sequences 
has been analyzed by Khodjasteh and Lidar \cite{khodj07}.

In the context of NMR the tuning of pulses to improve their properties
is well-known \cite{vande04}, mostly going back 
to the early work by Tycko \cite{tycko83}.
Static effects are compensated, yet no baths with internal dynamic
are considered \cite{cummi00,cummi03,brown04}. 
Numerical investigations how to reduce the influence
of classical noise also exist \cite{motto06}.
The possibilities of tailored pulses for specific examples of  sets of 
interacting qubits were investigated numerically in Ref.\ \onlinecite{sengu05}.
We will motivate our approach by comparing to the pulses
found there, see Sect.\ \ref{motivation}.

In our previous work \cite{pasin08a} we addressed 
the issue of a general dynamic quantum bath  analytically 
by an expansion in the shortness of the pulse, i.e., in powers of $\tau_p$.
The zeroth order of such an expansion is the instantaneous pulse
corresponding to a $\delta$ function. In the subsequent orders
the main difficulty is the non-commutation of the Hamiltonian describing the
pulse and of the Hamiltonian describing the coupling to the 
environment. Our approach separates both contributions up to
corrections in $\tau_p$. Then tailoring the pulse is used to
make as many perturbative corrections vanish as possible.
This improves the quality of the real pulses signficantly
as has been illustrated by numerical investigations  \cite{karba08}.

Note that we do not aim at eliminating the coupling to the bath during
the pulse. This is left to the pulse sequence in which the optimized
real pulses are intended to replace the ideal pulses.

The real pulses considered in Refs.\ \onlinecite{pasin08a}
and \onlinecite{karba08}
consider rotations of the qubit around a 
fixed axis. In the present paper we generalize the analytic expansion
to the case where the rotation  takes place around an axis varying in time. 
Thereby we aim at two goals. The first one is to be able to
describe experimental pulses more realistically where the pulses
are never strictly around a fixed axis.
The second one is to examine whether certain ideal pulses can be
better approximated by real general rotations than by
real fixed-axis rotations. This means we attempt to relax
the no-go results previously proven for real fixed-axis rotations 
\cite{pasin08a}.

We will show, however, that the no-go statements also apply
to general rotations. 
This sends the clear message to experiment that more complicated
rotations do not need to be considered, at least not for
the single qubit gates under study.

The paper is organized as follows. First, we give a mo\-ti\-va\-tion for
our analytical approach by comparison to numerical results
in Sect.\ \ref{motivation}. In Sect.\ \ref{ansatz} 
we introduce the model  and the ans\"{a}tze we will implement.
Also the  relations between the pulse shape, the axis and 
the angle of rotation are shown.
In Sect.\ \ref{GenEqs} the relevant Schr\"{o}dinger
 equations are solved formally. 
In Sect.\ \ref{expansion} the expansion 
in the shortness of the pulse is presented
by which we arrive at the final form for the general corrections.
They are discussed in Sect.\ \ref{proof} where we prove 
that the general rotations suffer from the same limitations
found for the fixed-axis rotations \cite{pasin08a}.
The findings are concluded and summarized in Sect.\ \ref{Conclusion}.

\section{Motivation\label{motivation}}

If the total Hamiltonian $H_\text{tot}(t)$ comprising the qubits $H$
and the pulse $H_0(t)$ reads $H_\text{tot}(t)=H+H_0(t)$ our primary
goal is to make the time evolution $U_p(\tau_p,0)$ 
during the pulse of length $\tau_p$  as close as possible to
\begin{eqnarray}
\label{time-evol-tot}
U_p(\tau_p,0) &=& T\left\{ 
\exp\left[-i\int_0^{\tau_p}H_\text{tot}(t)dt\right]\right\}
\\
&\approx & \exp(-i(\tau_p-\tau_s)H) \hat P_\theta \exp(-i\tau_s H)
\end{eqnarray} 
where $T$ stands for the conventional time ordering and
$\hat P_\theta$ for the unitary operator of the ideal pulse
rotating by the angle $\theta$. It is assumed to occur at the instant
$\tau_s\in [0,\tau_p]$, cf.\ Ref.\ \onlinecite{pasin08a} for details.

The literature so far \cite{tycko83,cummi00,cummi03,brown04,sengu05} 
pursues the goal 
\begin{equation}
U_p(\tau_p,0) \approx  \hat P_\theta.
\end{equation}
Both goals coincide if we focus on $\pi$ pulses ($\theta=\pi$)
with $\tau_s=\tau_p/2$ and $H=\sum_j \lambda_j \sigma_z^{(j)}$.
Then our goal implies
\begin{subequations}
\begin{eqnarray}
U_p(\tau_p,0) &=&  \exp(-i(\tau_p/2)H) \hat P_\pi \exp(-i(\tau_p/2) H)
\quad
\\
&=& \hat P_\pi \exp(i(\tau_p/2)H) \exp(-i(\tau_p/2) H)
\\
&=& \hat P_\pi.
\end{eqnarray}
\end{subequations}
The analogous argument holds if every second qubit is flipped and there
are Ising couplings  between adjacent qubits, see, e.g., Ref.\ 
\onlinecite{sengu05}.

For such systems Sengupta and Pryadko \cite{sengu05}
have proposed fine-tuned symmetric pulses labelled $S_L$ and  
$Q_L$ with $L\in\{1,2\}$. 
The $S_L$ pulses make the linear order ${\cal O}(\tau_p)$
vanish in $U_p(\tau_p,0)$, 
the $Q_L$ pulses  also make the quadratic order ${\cal O}(\tau_p^2)$ vanish.
Additionally, the $2L-1$ first derivatives of the pulse amplitudes
are zero at the beginning ($t=0$) and at the end ($t=\tau_p$) of the pulse.
We verified that all four pulses make the linear corrections
vanish which we have computed analytically in 
Ref.\ \onlinecite{pasin08a}. As far as the second order corrections are 
concerned, we must distinguish 
between terms which involve only the coupling to the environment $A$ 
(see Eq.~\ref{hamilt_gen_bath} in the next section) and terms where also the 
bath Hamiltonian $H_b$ appears, namely in the commutator $[A,H_b]$.
According to the notation in Ref.\ \onlinecite{pasin08a} the former corresponds
to the coefficient $\eta_{23}$ while the latter to the coefficients  
$\eta_{21}$ and $\eta_{22}$.
The $Q_L$ pulses make the former coefficient $\eta_{23}$ 
vanish. Note that the other two possible second order terms, $\eta_{21}$ and 
$\eta_{22}$, do not occur in an Ising model
since the coupling $A$ to the qubits commutes with the bath  Hamiltonian
$H_b$. So there is no contradiction to our proven result that a $\pi$ pulse
cannot be corrected in second order because this finding relied on
the generic model with internal dynamics, i.e., with $[A,H_b]\neq 0$.

From these results we see that all but one Fourier coefficients
found numerically, see Table II in Ref.\ \onlinecite{sengu05},
are determined by the
analytic non-linear equations derived in Ref.\ \onlinecite{pasin08a}.
The numerics required the variation of 
the Runge-Kutta solution to a high-dimensional set of differential equations.
Thus we conclude that the analytic expansion helps to avoid 
an important part of  tedious numerics. This observation shall serve
as additional motivation for the generalization of our analytic approach
to general rotations which we present in the sequel.

\section{Ans\"{a}tze\label{ansatz}}

Let us consider the following Hamiltonian of the qubit and its environment
\begin{equation}
\label{hamilt_gen_bath} 
H=H_b + \lambda A \sigma_z,
\end{equation} 
with $H_b$ representing a generic bath and $A$ its coupling operator 
to the qubit. This Hamiltonian is not the most general one 
because one spin direction is singled out. But it is
applicable to all experimental situations where the time $T_1$ is
much longer than $T_2$. Generically, this will be the case wherever
there is a large energetic splitting between the level $\sigma_z=-1$
and the level $\sigma_z=1$. Then the Hamiltonian \eqref{hamilt_gen_bath}
is the effective Hamiltonian in the rotating-wave approximation.

The internal  energy scale (inverse time scale) of $H_b$ shall be denoted by 
$\omega_b$. It is a measure for the internal dynamics of the bath.
Analogously, $\lambda$ measures 
the strength of the coupling between the qubit and the bath. 

The Hamiltonian of the control pulse reads
\begin{equation}
\label{hamilt_gen_pulse} 
H_0(t)=\vec{\sigma}\cdot\vec{v}(t),
\end{equation} 
where $\vec{\sigma}$ is the vector of the Pauli matrices and 
$\vec{v}(t)=(v_x(t),v_y(t),v_z(t))$ is a vector which defines the 
shape of the pulse along the three spin directions. 
The axis of rotation at the instant $t$ is given by $\vec{v}(t)/|\vec{v}(t)|$.

We concentrate on the evolution of the total system 
$H_\text{tot}(t)=H+H_0(t)$ comprising
qubit and bath during the application of the pulse. 
The total time evolution is given by the time-ordered
exponential in Eq.\ \eqref{time-evol-tot}.
In the sequel, we follow the same ideas as in Ref.\ \onlinecite{pasin08a}. 

The goal is to expand around the ideal instantaneous pulse which
we take to be located at the instant $\tau_s\in[0,\tau_p]$. To this end, we
split the time evolution in its part before $\tau_s$, $U_p(\tau_s,0)$,
and its part after $\tau_s$, $U_p(\tau_p,\tau_s)$. 
For these two evolutions we use the following ans\"atze
\begin{subequations}
\label{ansatz0}
\begin{equation}
\label{ansatz1}
 U_p\left(\tau_s ,0\right)=U_1\left(\tau_s ,0\right) \ 
T\left\{ e^{-i\vec{\sigma}\cdot
\int_{0}^{\tau_s}\vec{v}(t)dt}\right\} \ e^{-i \tau_s H}
\end{equation}
\begin{equation}
\label{ansatz2}
U_p\left(\tau_p ,\tau_s\right)=e^{-i (\tau_p-\tau_s)H } \
T\left\{e^{-i\vec{\sigma}\cdot \int_{\tau_s}^{\tau_p}
 \vec{v}(t) dt}\right\} \ U_2\left(\tau_p ,\tau_s\right),
\end{equation} 
\end{subequations} 
where the time ordering is required because of the 
non-commutation of $H_0(t)$ with itself at different instants
 $[H_0(t_1),H_0(t_2)]\neq 0$. This fact marks the major difference
to our previous analysis \cite{pasin08a}.

Both $U_1$ and $U_2$ are seen as the corrections which are necessary in order 
to factorize the two  exponentials of system and pulse even though
they do not commute. The corrections $U_1$ and $U_2$ are determined from the 
Schr\"{o}dinger equation, see Sect.\ \ref{GenEqs}.

The time-ordered exponential in (\ref{ansatz0}) can be translated into an 
overall rotation around an unknown axis ${\hat a}(\tau)$ 
($|{\hat a}(\tau)|=1$) about an angle $\psi(\tau)$
\begin{subequations}
 \label{ansatz_rot}
\begin{eqnarray}
 \label{ansatz_rot_plus}
e^{-i\vec{\sigma}\cdot\hat{a}(\tau){\psi(\tau)}/2}& := &
T_+\left\{e^{-i\vec{\sigma}\cdot\int_{\tau_s}^{\tau}\vec{v}(t)dt}\right\} 
\mathtt{,\ \mathrm{if}\ \Delta\tau\geqslant 0}
\\ \label{ansatz_rot_minus}
&:= & T_-\left\{e^{-i\vec{\sigma}\cdot
\int_{\tau_s}^{\tau}\vec{v}(t)dt}\right\} \mathtt{,\ 
\mathrm{if}\ \Delta\tau<0}\quad
\\ \label{ansatz_rot_sign}
&=& T_{\mathrm{sign(\Delta \tau)}}\left\{e^{-i\vec{\sigma}\cdot
\int_{\tau_s}^{\tau}\vec{v}(t)dt}\right\},\ \forall \tau
\end{eqnarray}
\end{subequations} 
where $\Delta\tau := \tau-\tau_s$ with $\tau \in [0,\tau_p]$ and $T_{\pm}$ 
stands for the increasing or decreasing time-ordering, respectively. 
Note that upon inversion the following identity holds
\begin{equation}
\label{relation_Tplus_Tminus}
\left\{T_-\left(e^{-i\vec{\sigma}\cdot
\int_{\tau_s}^{\tau}\vec{v}(t)dt}\right)\right\}^\dagger =
T_+\left(e^{i\vec{\sigma}\cdot\int_{\tau_s}^{\tau}\vec{v}(t)dt}\right).
\end{equation} 

For the sake of brevity, we introduce 
\begin{equation}
\hat{p}(t):= \vec{\sigma}\cdot\hat{a}(t){\psi(t)}/{2},
\end{equation}
which is a scalar operator.

Of course, there is a well-defined relation between $\vec{v}(t)$ 
on the one hand and  $\hat{a}(t)$ and $\psi(t)$ on the other. From the 
definition \eqref{ansatz_rot} we know
\begin{equation}
\label{Schroed_Hpulse}
i\partial_t e^{-i\hat{p}(t)} =
H_0(t)e^{-i\hat{p}(t)}. 
\end{equation}
The rotation in spin space can be explicitly written as
\begin{equation}
\label{ansatz_rot1}
e^{-i\hat{p}(t)}= \cos({\psi(t)}/{2})-
i (\vec{\sigma}\cdot\hat{a}(t))\sin({\psi(t)}/{2})
\end{equation}
Its time derivative simply reads
\begin{eqnarray}
\nonumber
\partial_t e^{-i\hat{p}} &=&
\frac{\psi^\prime(t)}{2}\Big(-\sin({\psi(t)}/{2})-i\vec{\sigma}\cdot
\hat{a}(t)\cos({\psi(t)}/{2}) \Big)
\\
\label{ansatz_rot2}
&& -i \vec{\sigma}\cdot\hat{a}^\prime(t) \sin({\psi(t)}/{2}) .
\end{eqnarray}
Inserting \eqref{hamilt_gen_pulse} in \eqref{Schroed_Hpulse}, then
exploiting \eqref{ansatz_rot1} and
\begin{equation}
\label{sigma_2}
\vec{\sigma}\left(\vec{\sigma}\cdot
\vec{n}\right)=\vec{n}+i\left(\vec{n}\times\vec{\sigma}\right),
\end{equation}
which holds for any vector $\vec{n}$, yields an explicit expression for 
$\partial_t e^{-i\hat{p}(t)}$ linear in $\vec{v}(t)$.
Its comparison to \eqref{ansatz_rot2} yields
\begin{eqnarray}
\label{a_psi_vs_v_noSigma}\nonumber
2\vec{v}(t)&=&\psi^\prime(t)\hat{a}(t)+\hat{a}^\prime(t)\sin\psi(t)
\\
&-&(1-\cos\psi(t))\left(\hat{a}^\prime(t)\times\hat{a}(t)\right).
\end{eqnarray} 
Multiplication with $\hat{a}(t)$ yields the derivative of $\psi(t)$
\begin{equation}
\label{psi_vs_v_a}
\vec{v}(t)\cdot\hat{a}(t)={\psi^\prime(t)}/{2}.
\end{equation}
Eq.\ \eqref{a_psi_vs_v_noSigma} clearly determines $\vec{v}(t)$ from
given $\hat{a}(t)$ and $\psi(t)$. But it can also be used to
find $\psi(t)$ and $\hat a(t)$ from  $\vec{v}(t)$ by integration
which is the way one has to take from an experimentally given pulse 
to its theoretical description.

\section{General equations\label{GenEqs}}

First, we consider $U_p\left(\tau_p ,\tau_s\right)$. From the Schr\"{o}dinger 
equation
\begin{equation}
 \label{sch_eq}
i \partial_\tau U_p\left(\tau ,\tau_s\right)=(H+H_0(\tau))
U_p\left(\tau ,\tau_s\right),
\end{equation}
and the ansatz (\ref{ansatz2}) we obtain
\begin{eqnarray}
\label{sch_eq_1}
\nonumber
H_0(\tau)U_p(\tau,\tau_s)&=&e^{-iH\Delta\tau}H_0(t)e^{-i\hat p(\tau)}
U_2(\tau,\tau_s) 
\\ 
&+& ie^{-iH\Delta\tau} e^{-i\hat p(\tau)}\partial_\tau 
U_2(\tau,\tau_s).
\end{eqnarray}
For the time derivative of $U_2$ this equation implies 
\begin{equation}
\label{schr_eq_2} i \partial_\tau U_2(\tau,\tau_s)=F(\tau) U_2(\tau,\tau_s)
\end{equation} 
where
\begin{eqnarray}
\label{F_iniz} 
F(\tau)&:=& e^{i\hat p(\tau)}\left[ \tilde{H}_0(\tau)-
H_0(\tau)\right]e^{-i\hat p(\tau)}
\end{eqnarray} 
and $\tilde{H}_0(\tau)= e^{iH\Delta\tau}H_0(\tau) 
e^{-iH\Delta\tau}$. The formal solution to this Schr\"{o}dinger equation 
reads
\begin{equation}
\label{sch_eq_sol_U2} U_2(\tau_p,\tau_s)=T_+\left\{
\exp\left(-i\int_{\tau_s}^{\tau_p}F(t) dt\right)\right\},
\end{equation} 
where $\Delta t :=t-\tau_s$.

The analogous procedure is used to obtain $U_1$ starting from 
\begin{equation}
 \label{sch_eq_cong}
-i \partial_\tau U_p\left(\tau_s,\tau\right) =
U_p\left(\tau_s,\tau\right) (H+H_0(\tau)),
\end{equation} 
where  $\tau\in [0,\tau_s]$. Finally, one finds
\begin{equation}
\label{sch_eq_sol_U1} 
U_1(\tau_s,0)=T_+\left\{\exp\left(-i\int_{0}^{\tau_s}F(t)
 dt\right)\right\}.
\end{equation} 
The time-dependent operator $F(t)$ is the same as the one appearing in Eq.\
 (\ref{F_iniz}). Note that $F(t)=0$ if there is no coupling between the 
qubit and the bath ($\lambda=0$) because $\tilde H_0(t)=H_0$ holds in this
case. Hence we have
\begin{equation}
F(t) = {\cal O}(t \lambda).
\end{equation}

Finally, we combine both corrections $U_1$ 
and $U_2$ to one correction $U_F(\tau_p,0)$
\begin{subequations}
\begin{eqnarray}
 U_p(\tau_p,0) &=& U_p\left(\tau_p,\tau_s\right)U_p\left(\tau_s,0\right)\\
&=& e^{-i\Delta\tau_p H} e^{-i\hat p(\tau_p)} U_F(\tau_p,0)
e^{i\hat p(0)} e^{-i \tau_s H} \qquad
\label{tot_ev_op} 
\end{eqnarray}
\end{subequations}
where $\Delta\tau_p=\tau_p-\tau_s$ and
\begin{subequations}
\begin{eqnarray}
\label{U_F1}
U_F(\tau_p,0)& := &U_2\left(\tau_p,\tau_s\right)U_1\left(\tau_s,0\right)
\\ 
&=&
T_+ \!\! \left\{ e^{-i\int_{\tau_s}^{\tau_p} F(t) dt}\right\}
T_+ \!\! \left\{ e^{-i\int_{0}^{\tau_s} F(t) dt}\right\}\qquad
\label{U_F2}
\\ 
&=& T_+\left\{ e^{-i\int_{0}^{\tau_p}  F(t) dt}\right\}.
\label{U_F}
\end{eqnarray}
\end{subequations}
Note that the intervals for the integration variable $t$
in \eqref{U_F2} are such that the global time-ordering in
\eqref{U_F} does not introduce any change compared to \eqref{U_F2}.

If the total correction $U_F(\tau_p,0)$ equals the identity then
the operator $e^{-i\hat p(\tau_p)} e^{i\hat p(0)}$ occurs in the
middle of the right hand side of \eqref{tot_ev_op}. 
We require this factor to be equal to the desired ideal pulse $\hat P_\theta$
\begin{equation}
\label{global_rot}
e^{-i\hat p(\tau_p)} e^{i\hat p(0)} = \hat P_\theta,
\end{equation}
for instance $\Theta=\pi$ for a $\pi$ pulse. The choice of the global
axis of rotation is arbitrary in the $xy$ plane of spin directions
in view of the rotation symmetry around $\sigma_z$. Hence we may assume
$\hat P_\theta =e^{i \sigma_y \Theta/2}$.

In view of the above, the essential issue is the deviation of
$U_F(\tau_p,0)$ from the identity. Thus we study $F(t)$
and rewrite \eqref{F_iniz} 
\begin{equation}
\label{F_1} 
F(t)=e^{i\hat p(t)} \vec{v}(t) \cdot \Delta\vec{\sigma}(\Delta t)
\ e^{i\hat p(t)},
\end{equation} 
where (with $\Delta t :=t-\tau_s$)
\begin{equation}
\label{Deltasigma}
\Delta\vec{\sigma}(\Delta t):= e^{iH\Delta t}\vec{\sigma} 
e^{-iH\Delta t}-\vec{\sigma}.
\end{equation}

From \eqref{F_1} we see that transforms of 
$\vec{\sigma}$ play an important role. Hence we define for 
later use
\begin{equation}
\label{S_definition}
\vec{S}(t) := e^{i\hat{p}(t)}\vec{\sigma}e^{-i\hat{p}(t)}.
\end{equation}
This vector operator is a rotation of $\vec{\sigma}$ about the 
axis $\hat{a}$ by the angle $\psi$. Hence it can be also be written
as
\begin{equation}
 \label{S_rotation_matrix}
\vec{S}(t)=D_{{\hat a}}(\psi)\vec\sigma,
\end{equation} 
where $D_{{\hat a}}(\psi)$ is the $3\times 3$ dimensional
matrix describing the rotation about the axis $\hat{a}$ by the angle $\psi$. 
The time dependences of  $\hat{a}(t)$ and $\psi(t)$ are omitted
to lighten the notation. For completeness, we also give the 
explicit form
\begin{equation}
 \label{S_fin}
\vec{S}(t)=\vec{\sigma}\cos\psi+\hat{a}(\vec{\sigma}\cdot\hat{a})
(1-\cos\psi)+(\vec{\sigma}\times\hat{a})\sin\psi ,
\end{equation}
which can be found using relations \eqref{ansatz_rot1} and 
\eqref{sigma_2}.

We will see shortly that the $z$-component $S_z(t)$ is what we need to know.
Hence we calculate
\begin{subequations}
\begin{eqnarray}
 \label{S_vs_n}
 S_z(t) &=& \hat z \cdot D_{{\hat a}}(\psi)\vec\sigma
\\
&=& (D_{{\hat a}}(-\psi) \hat z) \cdot \vec\sigma
\\
&=& \hat n(t) \cdot\vec\sigma
\label{n_def}
\end{eqnarray} 
\end{subequations}
where $\hat z$ is the unit vector in $z$ direction.
We put all the time dependence in the conventional
unit vector $\hat n(t):= D_{{\hat a}}(-\psi) \hat z$ in $\mathbbm{R}^3$.
It will enable us to give a geometrical interpretation to the final 
equations.

Finally, we state which effect a pulse of angle $\theta$ exerts on
$\vec{S}(t)$.
We start from \eqref{global_rot} and \eqref{S_definition} which imply
\begin{subequations}
\begin{eqnarray}
\vec{S}(0) &=& 
e^{i\hat{p}(\tau_p)} \hat P_\theta \vec{\sigma} \hat P_\theta^\dagger
e^{-i\hat{p}(\tau_p)}
\\
&=& e^{i\hat{p}(\tau_p)}\left(D_{\hat y}(\theta) \vec{\sigma} \right)
e^{-i\hat{p}(\tau_p)}
\\
&=& D_{\hat y}(\theta) \vec{S}(\tau_p).
\end{eqnarray}
\end{subequations}
Hence, a $\theta$ pulse rotates $\vec{S}(\tau_p)$ about $\hat y$ 
by the angle $\theta$. For $\theta=\pi$ this implies
\begin{equation}
\label{pi_condition}
S_z(0)=-S_z(\tau_p)\quad  \Leftrightarrow \quad \hat n(0)=- \hat n(\tau_p).
\end{equation}
Note that for other angles the implications on the
vector $\hat n(t)$ can be much less trivial in general.

\section{Expansion in $\tau_p H$\label{expansion}}

We consider the case where the duration $\tau_p$ of the pulse is short. 
This means that our expansion parameters are $\tau_p \lambda$ and
$\tau_p \omega_b$, or in shorthand we expand in $\tau_p H$.

The vector operator $\Delta\vec{\sigma} (\Delta t)$ is expanded in 
a power series of $\Delta t$, cf.\ Ref.\  \onlinecite{pasin08a},
$\Delta\vec{\sigma}(\Delta t)=
\sum_{n=1}^\infty\frac{i^n}{n!}(\Delta t)^n [[H,\vec{\sigma}]]_n$, with the 
notation  $[[H,\vec{\sigma}]]_1=[H,\vec{\sigma}]$, 
$[[H,\vec{\sigma}]]_2=[H,[H,\vec{\sigma}]]$ and so on.
For our Hamiltonian (\ref{hamilt_gen_bath}) the first and second order are
\begin{eqnarray}
\Delta\vec{\sigma}(\Delta t)=&-&2\Delta t\left(\vec{\sigma}\times
\hat{z}\right) \lambda A 
\nonumber \\
&-&(\Delta t)^2\left(\lambda [H_b,A]\ \vec{\sigma}\times\hat{z}+2\lambda^2 A^2 
\vec{\sigma}_\bot\right)
\nonumber \\ 
&+&{\cal O}(\Delta t^3),
\label{delta_sigma} 
\end{eqnarray} 
where $\vec{\sigma}_\bot := (\sigma_x,\sigma_y,0)$.

The perturbative computation of $U_F$ requires to solve 
Eq.\ \eqref{U_F}. This can be done by average Hamiltonian theory
\cite{magnu54,haebe76} which requires integrations over $F(t)$
defined in \eqref{F_1}. We show that the occurring integrals can
be simplified by integration by parts. To this end, we write
terms containg $\vec{v}(t)$ as time derivatives.

Inserting the expansion \eqref{delta_sigma} in \eqref{F_1} the terms 
$e^{i\hat p(t)}\vec{v}\cdot(\vec{\sigma}\times\hat z)e^{-i\hat p(t)}
=e^{i\hat p(t)} \hat z \cdot(\vec{v}\times \vec{\sigma})e^{-i\hat p(t)}$ 
and $e^{i\hat p(t)}\vec{v}\cdot \vec{\sigma}_\bot e^{-i\hat p(t)}$ occur. 
The combination of (\ref{Schroed_Hpulse}) with 
its Hermitean conjugate and with \eqref{sigma_2} provides us with
\begin{equation}
\partial_t\vec{S}(t) =
2e^{i\hat{p}(t)}\left(\vec{v}(t)\times\vec{\sigma}\right)e^{-i\hat{p}(t)}
\end{equation}
which expresses  the first of the above terms concisely as
\begin{equation}
\partial_t S_z (t) = 
2e^{i\hat p(t)}\vec{v}\cdot(\vec{\sigma}\times\hat z)e^{-i\hat p(t)}.
\end{equation}
Note that the given value of the total angle of rotation $\theta$ implies
 that $\partial_t S_z (t) \propto 1/|\vec{v}| \propto 1/\tau_p$.

The second term is found from the combination of (\ref{Schroed_Hpulse}) with 
its Hermitean conjugate \emph{without} using \eqref{sigma_2} implying
\begin{equation}
\partial_tS_z(t) = i e^{i\hat p(t)}
[\vec{v}\cdot\vec{\sigma},\sigma_z] e^{-i\hat p(t)}
\end{equation}
whence
\begin{equation}
\label{term1}
S_z(t) \partial_t S_z(t) = i e^{i\hat p(t)}
\sigma_z [\vec{v}\cdot\vec{\sigma},\sigma_z] e^{-i\hat p(t)}.
\end{equation}
Exploiting the identity
\begin{subequations}
\begin{eqnarray}
2\vec{\sigma}_\bot &=& \vec{\sigma} - \sigma_z \vec{\sigma}\sigma_z
\\
&=& \sigma_z [\sigma_z,\vec{\sigma}]
\end{eqnarray}
\end{subequations}
we finally arrive at
\begin{equation}
\label{term2}
-i S_z(t) \partial_t S_z(t) =
2 e^{i\hat p(t)}\vec{v}\cdot \vec{\sigma}_\bot e^{-i\hat p(t)}.
\end{equation}

With the help of \eqref{term1} and \eqref{term2} the expansion of $F(t)$ reads
\begin{eqnarray}
\label{F_vs_S}
F(t)=&-&\lambda A \Delta t\partial_tS_z(t)-i(\Delta t)^2\lambda [H_b,A]
\partial_t S_z(t)
\nonumber \\
&-&i\lambda^2 A^2(\Delta t)^2S_z(t)\partial_tS_z(t)
+{\cal O}(\Delta t^3).
\end{eqnarray}
Only the $z$-component of $\vec{S}$ is important. This a direct 
consequence of the coupling between the qubit and the bath in the 
Hamiltonian (\ref{hamilt_gen_bath}).
In this context it is noteworthy that higher terms in the expansion of 
$\Delta\sigma(\Delta t)$ (\ref{delta_sigma})  depend also only on the
two terms $\vec{\sigma}\times\hat{z}$ and $\vec{\sigma}_\bot$. 
This implies that all orders of $F(t)$ are functions of 
$S_z(t)$ and  $\partial_t S_z(t)$.

To reach an expansion of $U_F$ in terms of $\tau_p H$ we first express
it by means of the Magnus expansion \cite{haebe76,magnu54} 
\begin{eqnarray}
\label{U_F_Magnus}
U_F(\tau_p,0)=\exp\left[-i\tau_p(F^{(1)}+F^{(2)}+F^{(3)}...)\right],
\end{eqnarray}  
where each term $\tau_p F^{(j)}$ is of the order of $(\tau_p F)^j$.
The leading term is the time-average 
$\tau_p F^{(1)}=\int_0^{\tau_p}F(t)dt$ while the next-leading term comprises 
the commutator of $F(t)$ with itself at different instants 
$\tau_p F^{(2)}=\frac{-i}{2\tau_p}
\int_0^{\tau_p}dt_1\int_0^{t_1}dt_2[F(t_1),F(t_2)]$.

Because $F(t)$ itself is given by a series in $\Delta t$, see Eq.\ 
\eqref{F_vs_S}, the expansion \eqref{U_F_Magnus} is not yet the
desired expansion in $\tau_p$
\begin{eqnarray}
\label{U_F_Magnus_eta}
U_F(\tau_p,0)=\exp\left[-i(\eta^{(1)}+\eta^{(2)}...)\right],
\end{eqnarray} 
where $\eta^{(j)}$ represents the contribution of the power of $(\tau_p H)^j$.
Inserting \eqref{F_vs_S} in \eqref{U_F_Magnus} and expanding again
in $\tau_p$ yields the linear term
\begin{equation}
  \label{eta_1_noInt}
\eta^{(1)}=-\lambda A\int_0^{\tau_p}\Delta t\ \partial_t S_z(t) dt
\end{equation}
and the quadratic term
\begin{eqnarray}
  \label{eta_2_noInt}
\eta^{(2)}&=&-i\lambda^2A^2\left[\int_0^{\tau_p}(\Delta t)^2S_z(t)
\partial_tS_z(t)dt\right.
\nonumber \\
&+&\left.\frac{1}{2}\int_0^{\tau_p}dt_1\Delta t_1\int_0^{t_1}dt_2
\Delta t_2\left[\partial_{t_1}S_z(t_1),\partial_{t_2}S_z(t_2)\right]\right]
\nonumber \\
&-&i\lambda[H_b,A]\int_0^{\tau_p}(\Delta t)^2\partial_tS_z(t)dt.
\end{eqnarray}
These relations can be integrated by parts yielding
\begin{equation}
\label{eta_1}
\eta^{(1)}=-\lambda A\left[ [\Delta t S_z(t)]_0^{\tau_p}- \Sigma\right]
\end{equation} 
where we use the shorthand $\Sigma :=\int_0^{\tau_p}S_z(t)dt$.
The quadratic order reads 
\begin{equation}
\eta^{(2)} = -i\lambda  [H_b,A] \eta^{(2a)}
-({i}/{2})\lambda^2 A^2 \eta^{(2b)}
\end{equation}
where
\begin{subequations}
\label{eta_2}
\begin{eqnarray}
\label{eta_2a}
\eta^{(2a)} &=& \left[(\Delta t)^2S_z(t)\right]_0^{\tau_p}-2\int_0^{\tau_p}
\Delta t  S_z(t)dt
\\
\nonumber
\eta^{(2b)} &=&  \tau_s(\tau_p-\tau_s)
\left[S_z(\tau_p),S_z(0)\right]
\nonumber \\
&-&\left[(\tau_p-\tau_s)S_z(\tau_p)-\tau_sS_z(0), \Sigma \right]
\nonumber \\
&+& \int_0^{\tau_p}dt_1\int_0^{t1} dt_2 \left[S_z(t_1),S_z(t_2)\right].
\label{eta_2b}
\end{eqnarray}
\end{subequations}
In the simple case where the pulse acts only as a rotation in the $xz$ plane,
i.e., $v_x=v_z=0 \forall t$, one has $\hat a =\hat y$ and thus 
$S_z(t)= \sigma_z\cos\psi(t)+\sigma_x\sin\psi(t)$ according to \eqref{S_fin}.
With \eqref{psi_vs_v_a} one sees that  (\ref{eta_1}) and 
(\ref{eta_2}) reproduce the previous results obtained for fixed-axis
rotations \cite{pasin08a}.

\section{Discussion of the corrections\label{proof}}

In order to have pulses which well approximate ideal instantaneous pulses 
at $\tau=\tau_s$ we want to shape the real pulse such that
$\eta^{(1)}=\eta^{(2)}=0$.
The pulse shape is given in terms of the time dependences of
$\hat{a}$ and of $\psi$. They in turn determine the amplitude vector
 $\vec{v}$ uniquely via Eq.\ (\ref{a_psi_vs_v_noSigma}).

The conditions $\eta^{(1)}=\eta^{(2)}=0$ represent operator equations as they
stand. But they can be simplified using $\hat n(t)$ defined in \eqref{n_def}. 
Any equation linear in $S_z(t)=\hat n(t) \cdot\vec\sigma$
must hold for each vector component due to the linear independence
of the Pauli matrices. Hence $\eta^{(1)}=0$ is equivalent to
\begin{equation}
 \label{eta_1_n} 
(\tau_p-\tau_s)\hat n(\tau_p)+\tau_s\hat n(0)=
\int_0^{\tau_p}\hat n(t) dt.
\end{equation}
Equally, the vanishing of $\eta^{(2a)}$ is equivalent to 
\begin{equation}
\label{eta_2a_n}
(\tau_p-\tau_s)^2\hat n(\tau_p)-\tau_s^2\hat n(0) = 2\int_0^{\tau_p}\Delta t\ 
\hat n(t)\ dt.
\end{equation}
In case that $S_z(t)$ occurs quadratically (or in even higher powers) the
relation \eqref{sigma_2} helps to reduce the expression under study
to terms at most linear in $\vec{\sigma}$. 
Again, each component has to vanish separately
and we can thus transform $\eta^{(2b)}$ in \eqref{eta_2b} 
to a function of $\hat n(t)$ only.
Assuming that \eqref{eta_1_n}  holds the resulting expression equals 
\begin{equation}
\label{eta_2b_n}
\tau_s(\tau_p-\tau_s)\hat n(\tau_p)\times\hat n(0)=\int_0^{\tau_p}
 dt_1  \int_0^{t_1}dt_2\ \hat n(t_1)\times\hat n(t_2).
\end{equation}
It is very convenient that the complex condition on operators
 $\eta^{(1)}=\eta^{(2)}=0$ is simplified to three three-dimensional
vector equations \eqref{eta_1_n}, \eqref{eta_2a_n}, and \eqref{eta_2b_n}.
These three equations allow us to visualize the effect of the
generally rotating pulse. All one has to know about the pulse
is the orbit of $\hat n(t)$ on the unit sphere. This is the geometrical
interpretation of the corrections.

\subsection{No-go result for $\tau_s=\tau_p$} 

We pose the question whether a pulse can be tailored such
that $\tau_s=\tau_p$ holds. This would mean that a cleverly
designed pulse of finite duration corresponds to an ideal
instantaneous pulse at the very end of the real pulse.
Experimentally, this would be very advantageous because
one could start with measurements of the effects of such a
pulse without any delay, right after the end of the 
tailored pulse. 

But in Ref.\ \onlinecite{pasin08a} we proved
that such a pulse does not exist in the framework of fixed-axis
rotations. Hence we pose the question here again for a general
rotation. Unfortunately, the generalization of the pulse does
not help and we are able to prove even in the extended framework
that  $\tau_s=\tau_p$ is not possible.
For $\tau_p=\tau_s$ Eq (\ref{eta_1_n}) becomes
\begin{eqnarray}
 \label{ts_equal_tp}
\tau_p\hat n(0)-\int_0^{\tau_p}\hat n(t)\ dt=0.
\end{eqnarray} 
We multiply this equation by $\hat n(0)$ to reach
\begin{equation}
 \label{ts_equal_tp_2}
\tau_p=\int_0^{\tau_p}dt \cos\alpha(t)\leqslant\int_0^{\tau_p}dt=\tau_p,
\end{equation} 
where 
\begin{equation}
\label{cos_definition}
\cos\alpha(t):=\hat n(t)\cdot\hat n(0).
\end{equation}
The equality in \eqref{ts_equal_tp_2}  holds if and only 
if $\alpha(t)$ is a multiple of $2\pi$ almost everywhere in the
interval of integration. Hence only abrupt jumps of multiples of $2\pi$
comply with the condition \eqref{ts_equal_tp_2}. But such jumps
correspond to instantaneous pulses. Thus we conclude that $\tau_p=\tau_s$ is
impossible for real pulses in linear order and so to all orders.

This finding, independent from the total angle $\theta$, 
generalizes our previous no-go result from fixed-axis rotations to
pulses with varying axis of rotation.

\subsection{No-go result for second order corrections of $\pi$ pulses}

The ideal pulses with $\theta=\pi$ are the most important ones for
dynamical decoupling. We showed previously that they can be
approximated by real pulses with vanishing linear corrections
\cite{pasin08a,karba08}. But we proved for fixed-axis rotations
that it is impossible to tailor the $\pi$ pulse
such that the second order vanishes \cite{pasin08a}.
The proof holds if the decoherence bath possesses an internal dynamics,
i.e., $[A,H_b]\neq 0$, so that the prefactor of this term has to vanish.
Note that this is the decisive difference to the $Q_L$ pulses \cite{sengu05}
considered in Sect.\ \ref{motivation}.

Here we again pose the question whether $\pi$ pulses
can be corrected in second order if one uses general rotations.
Unfortunately, our finding is negative. The proof runs as follows.

For a $\pi$ pulse we know from \eqref{pi_condition} that
$\hat n(0)=- \hat n(\tau_p)$ holds. Next we multiply \eqref{eta_2a_n}
by $-\hat n(0)$ yielding
\begin{subequations}
\begin{eqnarray}
 \label{discussion_piPulse_2}
(\tau_p-\tau_s)^2+\tau_s^2&=& -2\int_0^{\tau_p}dt\ 
\Delta t\ \cos\alpha(t)
\\
&\leqslant& 2\int_0^{\tau_p}dt |\Delta t|
\\ 
&=&(\tau_p-\tau_s)^2+\tau_s^2.
\end{eqnarray}
\end{subequations}
The equality holds if and only if $\alpha(t) =\pi$ modulo $2\pi$ 
for $t>\tau_s$ and $\alpha(t) =0$ modulo $2\pi$
for $t<\tau_s$. So there must be at least one abrupt jump at $t=\tau_s$.
Hence only an instantaneous pulse satisfies  the second order condition
for a dynamical decoherence bath.

This finding generalizes our previous no-go result for $\pi$ pulses
from fixed-axis rotations to pulses with varying axis of rotation.

\section{Conclusions\label{Conclusion}}

In this work we have presented an analytical perturbative approach
to general short coherent control pulses for a two-level system, which may be
given by a $S=1/2$ spin or by a qubit. The spin is coupled (coupling
strength $\lambda$) to a quantum bath with internal dynamics (characteristic
frequency $\omega_b$). The small expansion parameter is the duration $\tau_p$ 
of the pulse, i.e., $\tau_p\lambda$ and $\tau_p\omega_b$ are taken to be
small. The starting point of the expansion is the ideal instantaneous pulse
which represents the zeroth order of the expansion with $\tau_p=0$.

We generalized the previous investigation of pulses which 
rely on rotations about a fixed axis \cite{pasin08a} to general rotation
about axes varying in time. The objective was twofold.

First, the general rotation eases the comparison with experiment
because it is  only approximately possible to realize
rotations about a fixed given axis. The generalized formalism
allows one to check the quality of complex rotations
by Eqs.\ \eqref{eta_1_n}, \eqref{eta_2a_n}, and
\eqref{eta_2b_n}. Moreover, these formulae render a geometric
interpretation possible which facilitates visualization.
The general pulse is characterized by
a path $\hat n(t)$ on a unit sphere.

Second, the generalized rotations allowed us to investigate
whether the previous no-go findings for fixed-axis rotations \cite{pasin08a}
can be circumvented by general rotation
about axes varying in time. But unfortunately, 
we proved that the general rotations comply with the same
limitations as the fixed-axis rotations:
(i) there is no real pulse which approximates an ideal
instantaneous pulse at the end of its time interval of finite duration.
(ii) $\pi$ pulses cannot be corrected in second order in $\tau_p$.
Though negative at first glance, the positive
message of this finding to experiment is that there
is no need in investigating complicated general rotations, at least
as far as the above limitations are concerned.

In which directions can the present results be extended?
Certainly, the Eqs.\ \eqref{eta_1_n}, \eqref{eta_2a_n}, and
\eqref{eta_2b_n} provide the basis for searching for
improved approximations for single quantum gates. For instance,
we could not find $\pi/2$ pulses, which realize the so-called Hadamard gate 
\cite{vande04}, with vanishing second order correction
among the fixed-axis rotations \cite{pasin08a}. We were not able
to prove the non-existence of such pulses. Thus we cannot exclude
the existence of a fixed-axis rotation approximating an ideal $\pi/2$ pulse
in second order correction. But the present generalized approach
definitely widens the range of pulses among which one can
look for such a well-approximating pulse.

Another direction of extension is to pass from the two-level system
to higher dimensional quantum systems coupled to a bath as they
occur in quantum optical manipulations.  The basic idea of our analytic 
approach is to disentangle the control pulse from the time evolution of
the system without external pulse. This idea will carry over to more
complex situations as well.

\begin{acknowledgments}
We are grateful to D.\ Lidar for bringing Ref.\ \cite{sengu05}
to our attention.
\end{acknowledgments}


\end{document}